\documentclass[acmtog]{acmart}

\author{Arjun Teh}
\email{ateh@andrew.cmu.edu}
\affiliation{%
  \institution{Carnegie Mellon University}
  \streetaddress{5000 Forbes Ave}
  \city{Pittsburgh}
  \state{PA}
  \postcode{15213}
  \country{USA}
}
\orcid{0000-0002-3579-2250}

\author{Delio Vicini}
\email{vicini@google.com}
\affiliation{%
  \institution{Google}
  \country{Switzerland}
}
\orcid{0000-0001-6330-6849}

\author{Bernd Bickel}
\email{berndbickel@google.com}
\affiliation{%
  \institution{Google}
  \country{Switzerland}
}
\orcid{0000-0001-6511-9385}

\author{Ioannis Gkioulekas}
\authornote{indicates equal contribution.}
\email{igkioule@andrew.cmu.edu}
\affiliation{%
  \institution{Carnegie Mellon University}
  \country{USA}
}
\orcid{0000-0001-6932-4642}

\author{Matthew O'Toole}
\authornotemark[1]
\email{motoole2@andrew.cmu.edu}
\affiliation{%
  \institution{Carnegie Mellon University}
  \country{USA}
}
\orcid{0000-0002-0740-9349}

\copyrightyear{2025}
\acmYear{2025}
\setcopyright{cc}
\setcctype{by}
\acmConference[SA Conference Papers '25]{SIGGRAPH Asia 2025 Conference Papers}{December 15--18, 2025}{Hong Kong, Hong Kong}
\acmBooktitle{SIGGRAPH Asia 2025 Conference Papers (SA Conference Papers '25), December 15--18, 2025, Hong Kong, Hong Kong}\acmDOI{10.1145/3757377.3763850}
\acmISBN{979-8-4007-2137-3/2025/12}

\usepackage{lens_mutations}

\acmSubmissionID{1250}
\begin{document}

\title{Automated design of compound lenses with discrete-continuous optimization}

\begin{abstract}
We introduce a method that automatically and jointly updates both continuous and discrete parameters of a compound lens design, to improve its performance in terms of sharpness, speed, or both. Previous methods for compound lens design use gradient-based optimization to update continuous parameters (e.g., curvature of individual lens elements) of a given lens topology, requiring extensive expert intervention to realize topology changes. By contrast, our method can \emph{additionally} optimize discrete parameters such as number and type (e.g., singlet or doublet) of lens elements. Our method achieves this capability by combining gradient-based optimization with a tailored Markov chain Monte Carlo sampling algorithm, using transdimensional mutation and paraxial projection operations for efficient global exploration. We show experimentally on a variety of lens design tasks that our method effectively explores an expanded design space of compound lenses, producing better designs than previous methods and pushing the envelope of speed-sharpness tradeoffs achievable by automated lens design.
\end{abstract}



\begin{teaserfigure}
	\includegraphics{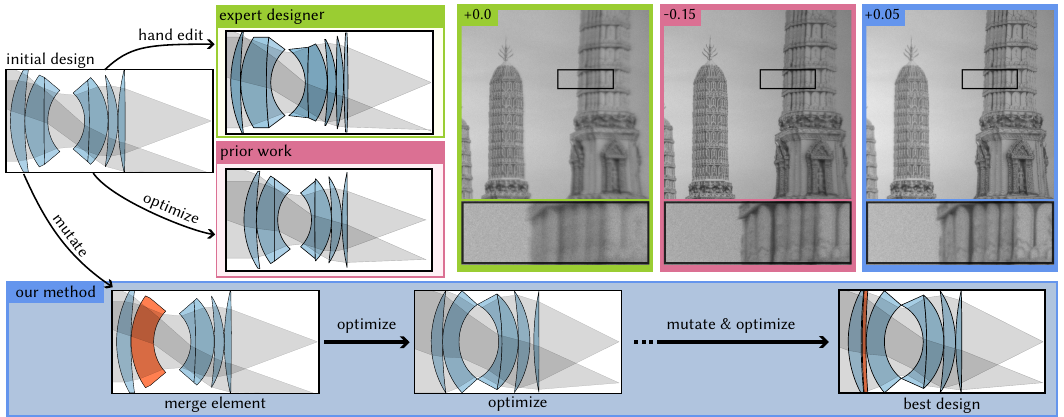}
	\caption{We develop a method that automatically explores the design space of compound lenses, by using Markov chain Monte Carlo sampling to combine gradient-based optimization with discrete changes to the number and type of lens elements. This combination allows our method to find designs that improve the sharpness and throughput of the initial lens design (in this example, the Nikon Nikkor-S 50mm $f/1.4$, released in 1962 \citep{reiley2014lensdesigns}), even after it has been optimized by prior gradient-based methods \citep{teh2024apertureaware}. Our method achieves image quality comparable to that of an improved lens designed by an expert (in this example, the Canon FD 50mm $f/1.2$, released in 1980). We report image brightness (top-left number of images) in terms of relative exposure.}
	\label{fig:teaser}
\end{teaserfigure}

\maketitle

\section{Introduction}

Modern lens design demands optimizing increasingly sophisticated compound lenses to meet increasingly challenging performance requirements. Though computational design tools exist, in practice these optimizations require close supervision by expert designers, who manually tune all aspects of the optical elements making up the compound lens. Obtaining good solutions requires tedious manual intervention, expert intuition, and trial and error.

Gradient-based optimization has become an essential part of the compound lens design process. Recent advances in differentiable rendering facilitate faster gradient-based optimization of continuous lens parameters (e.g., shapes of individual elements or distances between them)~\citep{wang2022differentiable,Tseng2021DeepCompoundOptics, sun2021endtoend,teh2024apertureaware}. Though such methods are invaluable for improving lens designs with a fixed \emph{topology} (number and type of lens elements), they cannot optimize the lens topology itself. It is up to the expert designer to perform discrete changes manually---strategically adding, changing, or removing elements from a base design---before handing the design back to the optimizer for further improvement. Critically, changing the lens topology is often the \emph{only} option available for meeting stringent performance requirements, for example: maintaining sharpness across the field of view when transitioning a lens from a half-frame to a full-frame sensor; increasing speed when designing a lens for extreme low-light conditions; or maintaining sharpness and speed when adapting a lens to smaller form factors.

Another way to assist expert designers is by generating good starting points. Methods using deep learning or genetic algorithms \citep{hoschel2018genetic, cote2021autolens,zoric2025combined} can help seed the design process, but the designs they output are often suboptimal: though they can make discrete changes to lens topology, they cannot effectively optimize continuous lens parameters, and thus cannot accurately assess the potential of different lens topologies. As a result, pushing the envelope in lens performance still requires expert designers to manually explore edits and iterate between discrete and continuous optimization.

These considerations highlight a critical gap in automated design of compound lenses: existing methods optimize only continuous or only discrete parameters of a system whose performance critically depends on \emph{joint} optimization of both types of parameters. We address this gap in automated lens design capabilities by developing a method that performs mixed discrete and continuous lens optimization automatically. Our method uses \emph{Markov chain Monte Carlo} (MCMC) sampling to combine gradient-based optimization of continuous lens parameters with \emph{transdimensional mutations} that alter the number of lens elements (\cref{fig:teaser}). We make such a combination practical through two core contributions:
\begin{itemize}[leftmargin=*]
	\item A sampling algorithm that facilitates mixing gradient updates and mutations without sacrificing optimization performance. 
	\item A set of mutations that use a projection operation to propose lenses with varied topologies that nevertheless remain paraxially equivalent to an original design.
\end{itemize}
These contributions enable our method to automatically explore a much larger design space of compound lenses than previously possible. We show experimentally that our method finds lenses with improved sharpness and speed compared to prior work that uses only gradient-based optimization. We provide interactive visualizations and an open-source implementation on the project website.%
\footnote{{\url{https://imaging.cs.cmu.edu/automated_lens_design}}}

\section{Related work}

\paragraph{Geometric optics design}
Lens designers commonly use computational tools~\cite{zemax, codev} to automatically optimize continuous lens parameters (e.g., curvature, size, and placement of lens elements). However, they typically have to rely on manual tuning and intuition to update the lens topology (e.g., number and type of lens elements) \citep{smith2008modern}. \citet{pedrotti2017introduction} and \citet{born2013principles} provide introductions to optics and lens design.

\paragraph{Differentiable rendering for lens design}
Differentiable rendering enables solving inverse rendering problems using gradient-based optimization~\cite{jakob2022drjit, Li:2018:DMC,zhang2022path}. Previous work focused on differentiating visibility discontinuities~\cite{Bangaru2022NeuralSDFReparam,Vicini2022sdf,Cai:2022:PSDR-MeshSDF, zhou2024path, wang2024simple} and reducing memory usage~\cite{vicini2021path, teh2022adjoint}. More recently, differentiable ray tracing methods have gained popularity for optimization of complex optical systems \citep{wang2022differentiable,teh2022adjoint,Tseng2021DeepCompoundOptics,sun2021endtoend}, often end-to-end with post-processing algorithms \citep{yang2024curriculum, cote2023differentiable}. \citet{cote2023differentiable} also extend gradient-based methods to optimize discrete material selection. Compared to prior work, our method uses \emph{aperture-aware differentiable ray tracing} \citep{teh2024apertureaware} for gradient-based optimization, but explores a larger design space by also modifying the number and type of lens elements.

\paragraph{Lens design exploration}
The design space of lenses is highly nonconvex and contains local minima. Prior work uses simulated annealing \citep{zoric2024global}, genetic algorithms \citep{hoschel2018genetic}, or large language models \citep{zoric2025combined} to mitigate these issues. These methods can help find performant designs, but use a predetermined number of elements. To expand the search space, \citet{betensky1993postmodern} proposed exchanging parts of a design with paraxially equivalent designs---a strategy our method adopts for mutations. Alternatively, \citet{sun2015lens} search over a library of off-the-shelf elements to generate designs without optimization. \citet{cote2021autolens} use deep learning to generate initial designs for subsequent optimization. Our method combines continuous optimization and discrete mutations, without requiring learning.

\paragraph{MCMC in computer graphics} Prior work in computational design often uses MCMC methods for discrete optimization problems. For example, \citet{yeh2012synthesizing} use reversible-jump MCMC to decide the placement of furniture in virtual rooms. \citet{desai2018assembly} use MCMC to optimize the placement of components of a mechanical assembly. More recently, \citet{barda2023sheetmetal} use MCMC to optimize the design of sheet metal parts. MCMC also finds application in rendering to estimate light transport integrals~\cite{veach1997metropolis}. Related to our work are methods that mix sampling distributions of varying dimensionality~\cite{bitterli2017reversible, otsu2017fusing}, or use gradients to guide sampling~\cite{luan2020langevin, li2015anisotropic}.

\paragraph{Quasi-stationary Monte Carlo}
Metropolis-Hastings adjusted processes reject samples, which can be inefficient, especially when each sample is costly to generate (e.g., in optimization). Quasi-stationary Monte Carlo (QSMC) methods sample from a target distribution without rejections. Unlike Metropolis-Hastings, QSMC methods do not require reversible transitions or detailed balance for convergence. \citet{pollock2020quasi} and \citet{wang2021regeneration} proposed QSMC algorithms that sample from a target distribution using Brownian motion. In rendering, \citet{holl2024jump} use the \emph{jump \restore algorithm}~\cite{wang2021regeneration} to mix global and local dynamics in Metropolis light transport. We use the jump \restore framework to develop a sampler that navigates the design space of lenses.

\section{Problem setup}\label{sec:setup}

We focus on compound lens design under four assumptions:
\begin{enumerate}[leftmargin=*]
	\item We assume each compound lens is a collection of refractive singlets or cemented doublets (\emph{lens elements}) that have spherical surfaces and are radially symmetric around the optical axis. 
	\item We do not include an aperture stop in the lens, thus the refractive elements determine the pupils and speed of the compound lens.
	\item We use geometric optics and ignore wave effects (e.g., diffraction). 
	\item We assume sequential lens design, where rays entering the lens transmit through each refractive surface once and in order from lens entrance to sensor; we thus ignore geometric effects where rays violate this assumption (e.g., interreflections, glare).
\end{enumerate}
With these assumptions, we represent a compound lens as a vector $\params \in \R^N$ that includes four parameters for each refractive surface, ordered from lens entrance to sensor: curvature, lateral extent, distance to the next surface, and refractive index after the surface. Thus, a compound lens with $K$ refractive surfaces has $N = 4 K$ parameters. 
%

We assume the compound lens is imaging a \emph{target plane} to a \emph{sensor plane}, both orthogonal to the optical axis at locations before and after the first and last (resp.) refractive surfaces. We simulate the image formation process using \emph{sequential ray tracing}: Given a ray $\ray \coloneq \paren{\pos_0, \vel_0}$ starting on the target plane at position $\pos_0 \in \R^3$ and direction $\vel_0 \in \Sph^2$, we propagate it through the lens with a sequence of refraction and propagation operations. Ray tracing ends when the ray either hits a stop (e.g., lens housing), or reaches the sensor plane at a position we denote $\trc\paren{\ray; \params} \in \R^3$ or $\trc\paren{\pos_0, \vel_0; \params} \in \R^3$, depending on context. We call rays that reach the sensor \emph{valid rays} and denote their set $\rayspace\paren{\params}\subset \R^3\times \Sph^2$. The function $\trc\paren{\cdot; \params}$ is differentiable with respect to $\params$, and its value and derivatives can be computed efficiently through \emph{primal} and \emph{adjoint} (resp.) sequential ray tracing \citep{teh2024apertureaware}, which we use as part of our method.


\paragraph{Design objectives.} We optimize $\params$ for a combination of losses that emphasize sharpness, speed, and conformity to focal length specifications. We define each loss for rays starting from a target plane location $\pos_0$, then aggregate for many such locations (\cref{eqn:total_loss}).
\begin{enumerate}[leftmargin=*]
	\item \emph{Sharpness:} We use a variance-based \emph{spot size loss}:
	\begin{align}
		\lossfun_{\text{spot}}(\pos_0; \params) &\coloneq \int_{\paren{\pos_0, \vel_0} \in \rayspace\paren{\params}} \norm{\trc\paren{\pos_0, \vel_0; \params} - \Tilde{\pos}(\pos_0)}^2 \ud \vel_0, \\
		\Tilde{\pos}(\pos_0) &\coloneq \frac{\int_{\paren{\pos_0, \vel_0} \in \rayspace\paren{\params}} \trc\paren{\pos_0, \vel_0; \params} \ud \vel_0 }{\int_{\paren{\pos_0, \vel_0} \in \rayspace\paren{\params}} \ud \vel_0}.
	\end{align}

	\item \emph{Speed:} We use a \emph{throughput loss} inspired by \citet{teh2024apertureaware}:
	\begin{equation}
		\lossfun_{\text{throughput}}(\pos_0; \params) \coloneq 1 - \frac{1}{T_0} \int_{\paren{\pos_0, \vel_0} \in \rayspace\paren{\params}} \ud \vel_0, \\ 
	\end{equation}
	where \(T_0\) is the maximum throughput determined by the maximum size of the design, which we specify as a hyperparameter. 

	\item \emph{Focal length:} We use a loss based on the target focal length \(f\), 
	\begin{equation}
		\lossfun_{\text{focal}}(\pos_0; \params) \coloneq \norm{\Tilde{\pos}(\pos_0) - \pos_{\mathrm{thin}}(\pos_0; f)}^2,
	\end{equation}
	where $\pos_{\mathrm{thin}}(\pos_0; f)$ is the sensor point predicted for $\pos_0$ from the Gaussian lens formula for a thin lens of focal length $f$.
\end{enumerate}
We include an additional regularization term that prevents lens elements from becoming thinner than a hyperparameter \(d_{\text{min}}\):
\begin{equation}\label{eqn:thickness}
	\lossfun_{\text{thickness}}(\params) \coloneq \sum\nolimits_{k=1}^K \max\left(d_{\text{min}} - t_{k}, 0\right)^2, 
\end{equation}
where \(t_{k}\) is the thickness of the \(k\)-th lens element.

Our total loss is a weighted combination of these losses, aggregated over multiple target locations where appropriate:
\begin{align}\label{eqn:total_loss}
	\lossfun(\params) \!\coloneq\!
	\sum\nolimits_{m=1}^M \!\bigl(&w_{\text{spot}} \lossfun_{\text{spot}}(\pos_0^m; \params) + w_{\text{throughput}} \lossfun_{\text{throughput}}(\pos_0^m; \params) \nonumber \\
	+ &w_{\text{focal}} \lossfun_{\text{focal}}(\pos_0^m; \params)\bigr) + w_{\text{thickness}} \lossfun_{\text{thickness}}(\params).
\end{align}

\subsection{Design with Markov chain Monte Carlo sampling}\label{sec:mcmc}

When the number $K$ of lens elements (and thus length $N$ of $\params$) is fixed a priori, the loss of \cref{eqn:total_loss} is differentiable using adjoint sequential ray tracing \citep{teh2024apertureaware}. Therefore, for fixed $N$, the design of a compound lens can be done using gradient-based optimization, an approach that has been popular in recent work \citep{teh2024apertureaware,wang2022differentiable,Tseng2021DeepCompoundOptics,sun2021endtoend}.

Unlike this prior work, we are interested in methods that design \emph{all aspects} of a compound lens, including both the discrete-valued number $K$ of elements and their continuous-valued parameters $\params$. Though $K$ is not amenable to gradient-based optimization, we would like any such method to continue using gradient-based optimization for $\params$. We enable \emph{mixed discrete-continuous design} by treating it as a sampling problem that we attack with \emph{Markov chain Monte Carlo} (MCMC) algorithms. To this end, we first convert the loss in \cref{eqn:total_loss} into a positive Boltzmann distribution to be \emph{maximized}:
\begin{equation}\label{eqn:boltzmann}
	\targetdist(\params) \coloneq e^{-\frac{\lossfun(\params)}{\temperature}},
\end{equation}
where $\temperature$ is a \emph{temperature} that controls the ``sharpness'' of the distribution: higher $\temperature$ allows exploring larger regions of the design space, whereas lower $\temperature$ forces exploration of only high-value regions. We treat $\temperature$ as a fixed hyperparameter, though future work could explore \emph{simulated annealing} methods that progressively lower temperature. 

\paragraph{Metropolis-Hastings and its pitfalls.} The most popular MCMC method is the \emph{Metropolis-Hastings} (MH) algorithm \citep{hastings1970}. Given a lens $\params$, it uses a \emph{proposal distribution} $\pprop$ to sample a lens proposal $\params' \sim \pprop(\params\to\params')$, and compute an acceptance probability:
\begin{equation}\label{eqn:mh}
	\alpha \coloneq \min\Biggl\{1,\, \exp\Biggl(\frac{\lossfun\paren{\params}-\lossfun\paren{\params'}}{T}\Biggr) \cdot \frac{\pprop\paren{\params'\to\params}}{\pprop\paren{\params\to\params'}}\Biggr\}.
\end{equation}
The proposal is randomly either accepted with probability $\alpha$ and used to update the lens ($\params \gets \params'$), or rejected leaving the lens unchanged. The exponential term in \cref{eqn:mh} favors accepting proposals $\params'$ that improve (reduce) $\lossfun$. Under certain conditions on $\pprop$, this algorithm will sample lenses proportionally to $\lossfun$.

The appeal of MH for mixed discrete-continuous design lies in the great flexibility it affords the user in selecting the proposal distribution $\pprop$. We can use a \emph{mixture} proposal distribution:
\begin{equation}\label{eqn:proposal}
	\pprop\paren{\params\to\params'} = \weight \ppert\paren{\params\to\params'} + \paren{1 - \weight} \pmuta\paren{\params\to\params'},
\end{equation}
which at random updates lens parameters $\params$ either continuously by sampling $\ppert$---a \emph{perturbation} that leaves the dimension $N$ the same---or discretely by sampling $\pmuta$---a \emph{mutation} that changes $N$. Moreover, for $\ppert$, we can use \emph{Langevin perturbations} $\params' \sim \Normal(\params - \eta\ \nicefrac{\dd \lossfun}{\dd \params}, \eta)$ \citep{roberts1996exp}, or equivalently:
\begin{equation}\label{eqn:langevin}
	\params' \gets \params - \eta \frac{\dd \lossfun}{\dd \params} + \sqrt{\eta}\cdot \noise,
\end{equation}
where $\noise$ is a vector of normal random variates, and the gradient term can be computed using adjoint sequential ray tracing. The Langevin perturbations in \cref{eqn:langevin} mimic gradient descent, except for the addition of noise, allowing MH to incorporate gradient-based optimization of continuous lens parameters. In practice, it is common to use adaptive step sizes $\eta$ in \cref{eqn:langevin} to improve sampling performance \citep{li2015anisotropic,luan2020langevin}, analogously to gradient descent algorithms such as Adam \citep{kingma2017adam}.

MH with mixed Langevin perturbations and discrete mutations has been successful elsewhere in graphics, notably in Monte Carlo rendering \citep{li2015anisotropic,luan2020langevin}. Unfortunately, we have empirically found it difficult to use this approach for mixed discrete-continuous design of compound lenses, for two reasons:
\begin{enumerate}[leftmargin=*]
	\item The sharpness and speed of a compound lens $\params$ with many elements (e.g., $K \ge 4$) is \emph{very} sensitive to random perturbations. Langevin perturbations include noise $\varepsilon$ (\cref{eqn:langevin}), and thus almost always produce badly performing lenses that are rejected. Preventing this behavior requires keeping the noise variance---and thus step size $\eta$---very small. The result in either case is that optimization requires prohibitively many sampling iterations.
	\item Lens mutations that have high acceptance probability generally require refining a lens after increasing or decreasing its elements. Such post-mutation refinement requires continuous optimization, which makes it challenging---or impossible---to compute the reverse proposal distribution $\pprop(\params^\prime \to \params)$ in \cref{eqn:mh}.
\end{enumerate}
We provide experimental evidence for both issues in \cref{sec:results}. These issues arise from the requirement for \emph{reversible proposals} in MH, which necessitates the addition of noise in \cref{eqn:langevin} and the computation of $\pprop(\params^\prime \to \params)$ in \cref{eqn:mh}. In \cref{sec:restore}, we address both issues by using a different MCMC algorithm that lifts the reversibility requirement alongside providing other advantages. Then, in \cref{sec:mutations}, we design effective discrete lens mutations, leveraging the flexibility from no longer being constrained by reversibility. 

\begin{algorithm}[t]
	\caption{\Proc{SingleRestoreStep}$(\params, R, \gamma,  \targetdist, \cc)$}
	\label{alg:single_restore_step}
	\begin{algorithmic}[1]
		\algblockdefx[Name]{Struct}{EndStruct}
		[1][Unknown]{\textbf{struct} #1}
		{}
	\algtext*{EndStruct}
	\algblockdefx[Name]{FORDO}{ENDFORDO}
		[1][Unknown]{\textbf{for} #1 \textbf{do}}
		{}
	\algtext*{ENDFORDO}
	\algblockdefx[Name]{WHILEDO}{ENDWHILEDO}
		[1][Unknown]{\textbf{while} #1 \textbf{do}}
		{}
	\algtext*{ENDWHILEDO}
	\algblockdefx[Name]{IF}{ENDIF}
		[1][Unknown]{\textbf{if} #1 \textbf{then}}
		{}
	\algtext*{ENDIF}
	\algblockdefx[Name]{ELSE}{ENDELSE}
		{\textbf{else}}
		{}
	\algtext*{ENDELSE}
	\algblockdefx[Name]{IFTHEN}{ENDIFTHEN}
		[2][Unknown]{\textbf{if} #1 \textbf{then} #2}
		{}
	\algtext*{ENDIFTHEN}
	\algblockdefx[Name]{RETURN}{ENDRETURN}
		[1][Unknown]{\textbf{return} #1}
		{}
	\algtext*{ENDRETURN}
	\algblockdefx[Name]{COMMENT}{ENDCOMMENT}
		[1][Unknown]{\textcolor{commentcolor}{\(\triangleright\) \textit{#1}}}
		{}
	\algtext*{ENDCOMMENT}

	\Require \textit{A compound lens $\params$, a reservoir $R$, a weight parameter $\gamma$, a target distribution $\targetdist$, a constant $\cc$.}
 	\Ensure \textit{A new compound lens $\Tilde{\params}$, an updated reservoir $R$.}
		\COMMENT[Perform a gradient step]\ENDCOMMENT
		\State $\Tilde{\params} \gets \Proc{GradientStep}(\targetdist, \params)$
	 	\COMMENT[Compute the termination probability]\ENDCOMMENT
		\State $\regenrate \gets \frac{\targetdist(\params)-\targetdist(\tilde{\params})+\cc}{\targetdist(\params)+\cc}$
		\COMMENT[Check whether to terminate perturbation sequence]\ENDCOMMENT
		\IFTHEN[$\Proc{SampleUniform}\bracket{0,1} < \regenrate$]{
			\COMMENT[Update the reservoir with the current lens]\ENDCOMMENT
			\State $R \gets \Proc{UpdateReservoir}(R, \Tilde{\params})$
			\COMMENT[Sample and mutate new lens]\ENDCOMMENT
			\State \textbf{if} $\Proc{SampleUniform}\bracket{0,1} < \gamma$ \textbf{then}
				\State $\qquad\Tilde{\params} \gets \Proc{SampleGlobal}()$
			\State \textbf{else}
				\State $\qquad\Tilde{\params} \gets \Proc{SampleReservoir}(R)$
			\State $\Tilde{\params} \gets \Proc{MutateLens}(\Tilde{\params})$
		}\ENDIFTHEN
		\COMMENT[Return new lens and updated reservoir]\ENDCOMMENT
		\RETURN[$\Tilde{\params}, R$]\ENDRETURN
	\end{algorithmic}
\end{algorithm}

\section{The \restore algorithm}\label{sec:restore}

We propose to use an MCMC algorithm based on so-called \emph{\underline{r}andomly \underline{e}xploring and \underline{sto}chastically \underline{re}generating} (\restore) processes \citep{wang2021regeneration}---we use the name \restore also for the algorithm associated with these processes. \restore uses two proposal distributions, one for small perturbations, and another for large changes. The \restore literature refers to these distributions as \emph{local dynamics} and \emph{regeneration probability} (resp.); for clarity, we continue to use the terms \emph{perturbation} and \emph{mutation} (resp.) from \cref{sec:mcmc}.

Starting from an initial lens $\params$, \restore performs a sequence of perturbations updating $\params$ (``local dynamics simulation''). At each perturbation, it uses $\params$ to compute a \emph{termination probability} to randomly decide whether to continue perturbations or terminate the sequence. Upon termination, it updates $\params$ using the mutation distribution (``regeneration''), then restarts a perturbation sequence. Compared to MH, \restore has no rejections, and does not require reversible perturbation and mutation distributions---both differences that benefit optimization \citep{holl2024jump}. Instead of reversibility, \restore relies on careful selection of the mutation distribution and termination probability to ensure it samples the target distribution $\targetdist$ \citep{wang2021regeneration}. We first detail our choices for the perturbation and mutation distributions, then explain how to determine the termination probability; along the way, we elaborate on advantages over MH. We summarize our version of \restore in \cref{alg:single_restore_step}.


\paragraph{Perturbation distribution.} As \restore does not require reversible proposals, we use gradient perturbations \emph{without noise}:
\begin{equation}\label{eqn:gradient}
	\params \gets \params - \eta \frac{\dd \lossfun}{\dd \params}.
\end{equation}
Thus, a perturbation sequence becomes equivalent to gradient descent. 
The lack of noise is crucial for performance: Whereas in \cref{eqn:langevin} the noise term required keeping step size $\eta$ small, in \cref{eqn:gradient} we can use large step sizes to accelerate optimization. We use Adam \citep{kingma2017adam} to determine adaptive step sizes from the history of the current sequence. As we explain below, the sequence terminates randomly with a probability that adapts to how much progress gradient steps make towards improving $\params$.

\begin{figure}
	\includegraphics{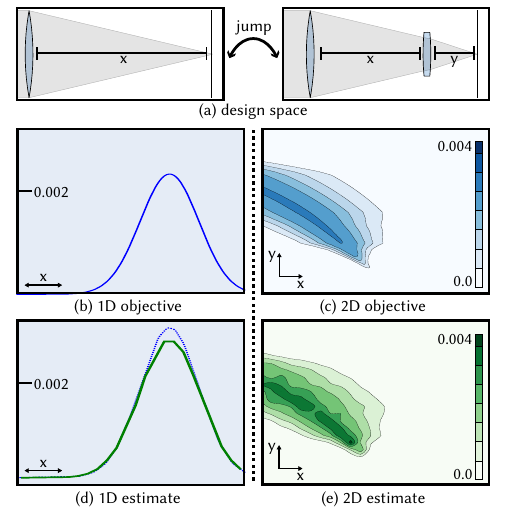}
	\caption{We construct a toy lens design problem (a) to validate the sampling accuracy of our method. The sampler can choose between optimizing the distance of a singlet from the sensor (1D target distribution, left), or the distances between two singlets and the sensor (2D target distribution, right). Even with an approximate termination probability, our method (d--e) correctly samples from the target distributions (b--c) in both 1D and 2D.}
	\label{fig:restore_demo}
\end{figure}

\paragraph{Mutation distribution.} \restore requires that the mutation distribution can sample any lens $\params$ with non-zero probability. In practice, the choice of mutation distribution is further constrained by its role in determining the termination probability---through an integral operator relationship that is difficult to compute analytically except for trivial distributions (e.g., uniform) \citep[Section 3]{wang2021regeneration}. We follow \citet{pollock2020quasi}, who show that using a mutation distribution that selects from a reservoir of previous samples allows a tractable approximation of the termination probability.

In particular, we first sample a lens $\params$ from a mixture distribution
\begin{equation}\label{eqn:mutation}
	\gamma \pglobal\paren{\params} + \paren{1-\gamma} \pres\paren{\params, R},
\end{equation}
where:
\begin{enumerate*}
	\item $\pglobal$ samples $\params$ entries independently and uniformly within some upper and lower bounds;
	\item $\pres$ samples $\params$ from a reservoir $R$ of previously samples, as we explain below.
\end{enumerate*}
We set the hyperparameter $\gamma$ to a small value, to ensure that the mutation distribution satisfies the \restore requirement while mostly sampling from the reservoir. After sampling $\params$, we mutate it as in \cref{sec:mutations}. 

We populate $R$ by storing the top $N$ lenses (in terms of $\targetdist$) sampled upon perturbation chain terminations. The reservoir effectively allows ``backtracking'' to a previous high-performing lens when a perturbation chain gets trapped exploring a bad region of the design space. Without this backtracking, the mutation distribution would be extremely unlikely to find a reasonable lens by randomly sampling the design space. Using the reservoir allows ``warm starting'' a new perturbation chain from a previous reasonable lens.

\paragraph{Termination probability} \Citet{pollock2020quasi} show that using gradient steps for perturbations and reservoir sampling for mutations simplifies the termination probability to:
\begin{equation}\label{eqn:regenrate}
	\regenrate \gets \frac{\targetdist(\params)-\targetdist(\tilde{\params})+\cc}{\targetdist(\params)+\cc},
\end{equation}
where $\cc > 0$ is a hyperparameter. Intuitively, $\regenrate$ increases, and thus termination becomes more likely, when the relative improvement of $\targetdist$ after a gradient step becomes small. Thus, \restore \emph{adaptively} determines to stop perturbations and use a mutation when gradient optimization gets stuck at a local maximum of $\targetdist$. Though \cref{eqn:regenrate} is an approximation, we show in \cref{fig:restore_demo} that using it in \restore still allows sampling from the target distribution $\targetdist$.

\section{Lens mutations}\label{sec:mutations}

\begin{figure}
	\includegraphics{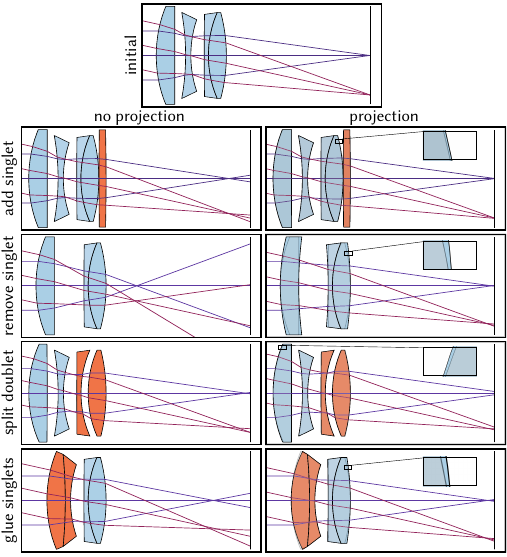}
	\caption{Left column: We consider four types of mutations of a compound lens. Applying such a mutation often leads to a lens with significantly worse performance than the original (diverging rays on the sensor plane). Right column: We alleviate this issue by using a \emph{paraxial projection} refinement process that updates the mutated lens to have the same first-order focusing behavior as the original lens. For visualization, we overlay the original lens in grey over the refined one. Even though paraxial projection makes only small changes to lens elements, it greatly improves performance (converging rays on the sensor plane), while being very efficient to run.}
	\label{fig:projection_demo}
\end{figure}

We implement four mutations (\cref{fig:projection_demo}) inspired by strategies in lens design textbooks \citep{smith2008modern}: singlet addition, singlet removal, singlet gluing, and doublet splitting. \citeauthor{smith2008modern} also suggests strategies for choosing an element and mutation (e.g., splitting high-curvature singlets). However, for simplicity we use uniform element and mutation sampling, which empirically performed comparably to sampling schemes derived from these strategies. Following the mutation, we refine the lens using a \emph{paraxial projection} process, which we explain next using singlet addition as an example---the same process applies to the other mutations. 
We can perform this refinement because the overall mutation distribution does not need to be reversible.

If we simply add a singlet to the current compound lens, the resulting lens will almost certainly have drastically worse focusing and throughput performance than the original lens. Though RESTORE will still attempt to optimize the mutation proposal, optimization will likely get stuck fast, requiring a new mutation---and thus wasting the previous ones. We therefore need a way to quickly fine-tune the lens from a mutation proposal so that it performs closer to the original lens. We do so using \emph{ray transfer matrix analysis}---a paraxial approximation to ray tracing---as a proxy for \cref{eqn:total_loss}. We emphasize that we use the paraxial approximation only for this refinement process: our overall method uses full ray tracing to design lenses while accounting for non-paraxial effects (\cref{sec:setup}).

\begin{figure}
	\includegraphics{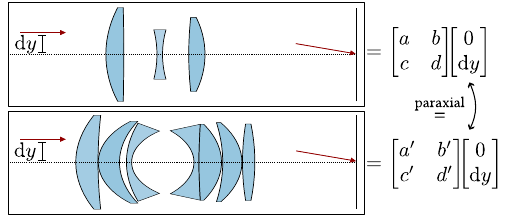}
	\caption{The $2 \times 2$ ray transfer matrix describes how a light ray parallel and close to the optical axis propagates through a compound lens. If two lenses have the same effect on such a ray, they are \emph{paraxially equivalent}.}
	\label{fig:ray_transfer_matrix}
\end{figure}

\paragraph{Paraxial lens optics.} With the paraxial approximation of radially symmetric optics, we parameterize a ray as the 2D vector of its distance $y$ and angular deviation $\phi$ from the optical axis. We also model propagation of the ray through a sequential compound lens as matrix-vector multiplication with $2\times 2$ \emph{ray transfer matrix} $\raytransfermat$:
\begin{equation}
	\begin{bmatrix} \phi_{\text{f}} \\ y_{\text{f}} \end{bmatrix} = \raytransfermat  \begin{bmatrix} \phi \\ y \end{bmatrix},
\end{equation}
We overview this matrix for compound lenses comprising spherical elements, and refer to \citet[Chapter 18]{pedrotti2017introduction} for details.

As a ray traveling through a lens undergoes a sequence of propagation and refraction operations, first we model each such operation as a ray transfer matrix. Propagation by a distance $d$ and refraction at a spherical surface of curvature $\kappa$ correspond to matrices:
\begin{equation}
\raytransfermat_{\mathrm{p}}\paren{d} \coloneq \begin{bmatrix} 
		1 & 0 \\
		d & 1
	\end{bmatrix},\quad \raytransfermat_{\mathrm{r}}\paren{\kappa, \eta_\mathrm{o}, \eta_\mathrm{i}} \coloneq \begin{bmatrix} 
		\nicefrac{\eta_\mathrm{i}}{\eta_\mathrm{o}} & \nicefrac{\kappa(\eta_\mathrm{i} - \eta_\mathrm{o})}{\eta_\mathrm{o}} \\
		0 & 1
	\end{bmatrix},
\end{equation}
where $\eta_\mathrm{i}$ and $\eta_\mathrm{o}$ are the refractive indices before and after refraction. 

We can then model a compound lens as the product of the refraction and propagation matrices corresponding to its elements. For example, a singlet is represented paraxially as $\raytransfermat_{\mathrm{r}} \raytransfermat_{\mathrm{p}} \raytransfermat_{\mathrm{r}} \raytransfermat_{\mathrm{p}}$, where the last matrix considers the distance from the final lens surface to the sensor plane. Given a compound lens $\params$, we denote its ray transfer matrix $\raytransfermat(\params)$, which can always be written in the form:
\begin{equation}
	\raytransfermat(\params) = \begin{bmatrix} a & -\nicefrac{1}{f} \\ b & c \end{bmatrix},
\end{equation}
where $f$ is the focal length, and $a$, $b$, $c$ are coefficients encoding other first-order properties of the compound lens. $\raytransfermat(\params)$ provides a fixed-sized abstraction of the compound lens, no matter its complexity (e.g., number $K$ of elements). Moreover, two compound lenses will have similar (up to first order) focusing behavior if their ray transfer matrices are equal, no matter how different their internal designs. We thus define the following notion of \emph{approximate} equivalence between two compound lenses---emphasizing focusing of rays parallel to the optical axis and at infinite conjugacy ($y = 1, \phi = 0$).

\begin{dfn}[label={def:paraxialeq}]{Paraxial equivalence}
	Two compound lenses \(\params_a\) and \(\params_b\) are \emph{paraxially equivalent} if \[\raytransfermat(\params_a)\begin{bmatrix} 0 \\ 1 \end{bmatrix} = \raytransfermat(\params_b) \begin{bmatrix} 0 \\ 1 \end{bmatrix}.\]
\end{dfn}


%

\paragraph{Paraxial projection} 
After adding a singlet to a lens $\params$, we refine the augmented lens $\params^*$ so that it is paraxially equivalent to $\params$. We formulate this refinement as a constrained optimization problem, which we solve using Newton's method and Lagrange multipliers:
\begin{equation}
    \min_{\params^*} ~~ \norm{\params - \params^*}^2 \quad \text{s.t.} \quad \paren{\raytransfermat(\params) - 
	\raytransfermat\paren{\params^*}} \begin{bmatrix} 0 \\ 1 \end{bmatrix} = 0. \label{eqn:projection}
\end{equation}
We keep entries of $\params^*$ for the added singlet fixed. 
Empirically, we found that this process, which we term \emph{paraxial projection}, results in much larger objective improvements than gradient descent with full ray tracing for equal runtime (30 iterations). The paraxial projection is locally unique and thus has a full-rank Jacobian of $\params^*$ with respect to $\params$ (\cref{sec:appendix_jacobian}). This property, though not needed by our method, allows using paraxial projection in MH-based samplers requiring reversibility, and we compare with such a sampler in \cref{sec:results}.

\paragraph{Comparison with automatic design search} Commercial software \citep{synopsys} includes \emph{automatic design search} procedures, which mutate compound lens designs through \emph{automatic lens insertion} (AEI) and \emph{automatic lens delete} (AED) methods, similar in spirit to our singlet-addition and singlet-removal mutations (resp.). AEI randomly adds singlets that initially have near-zero thickness, so that they do not impact the design. In our experiments, we found that such singlets remained thin even after several gradient iterations, and mitigating this issue required using expensive line search methods and a large weight on the thickness penalty loss (\cref{eqn:thickness}). AED removes elements by first attempting to make them have near-zero thickness through the addition of extra thickness penalties to the loss function. In our experiments, we found that AED was very sensitive to hyperparameter tuning, and significantly slowed down convergence. By contrast, our method supports a richer set of mutations that modify the design without causing gradient descent to get stuck or requiring dynamically changing the loss function.


\begin{figure*}[!t]
	\includegraphics{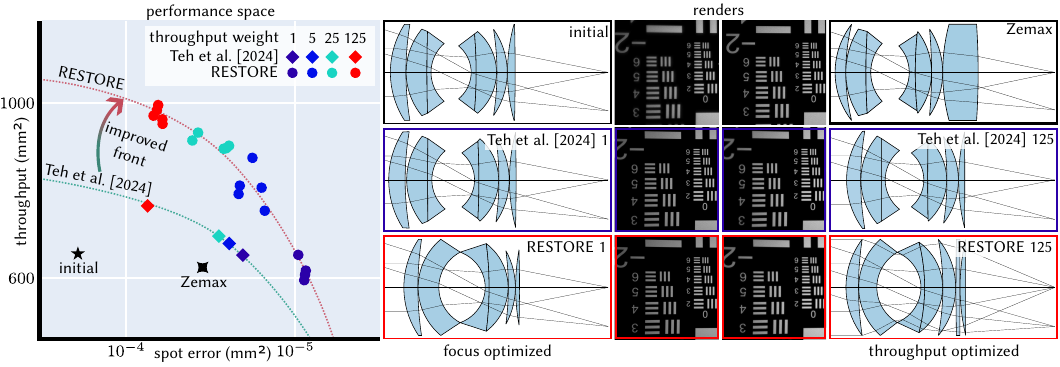}
	\caption{By varying the throughput and spot error weights in the optimization loss, we can explore designs that achieve different tradeoffs between lens speed and sharpness, tracing a Pareto front. Without mutations, lens designs are limited to the Pareto front determined by the initial design's topology. As our method can add and remove elements from the design, it is able to explore a larger space of designs and expand the Pareto front to achieve better tradeoffs.}
	\label{fig:pareto}
\end{figure*}

\section{Experiments}\label{sec:results}

We show results from experiments on several lens design tasks. The project website includes interactive visualizations and more results.

\paragraph{Implementation details.} We implemented our method in Python, using JAX \citep{jax2018github} for GPU acceleration, and Optimistix \citep{optimistix2024} for paraxial projection. To compute derivatives of \cref{eqn:total_loss}, we implemented aperture-aware differentiable ray tracing \citep{teh2024apertureaware}, which was previously shown to improve lens optimization compared to alternatives \citep{wang2022differentiable}. We assume a \qty{35}{\milli\meter} full-frame sensor, and optimize for four target plane positions in \cref{eqn:total_loss}, selected so that their thin-lens sensor-plane projections are uniformly distributed. We set the temperature $\temperature$ in \cref{eqn:boltzmann} so that $\targetdist\paren{\params_0} \approx 0.5$, the size of the reservoir $R$ to 5 and $\gamma = 0.02$ in \cref{eqn:mutation}, and $\cc = 2$ in \cref{eqn:regenrate}. Initial designs are from \citet{reiley2014lensdesigns}. On a workstation with an NVIDIA RTX 3090 GPU, our implementation performs 12,000 iterations in \qty{40}{\min}. We provide an open-source implementation of our method on the project website. When we use \citet{zemax}, we optimize the default merit function for spot error and added objectives for focal length, running Zemax's damped least squares until convergence.

\paragraph{Expanding the lens sharpness-speed Pareto front}
\Cref{fig:pareto} shows that our method allows exploring a larger tradeoff space of lens designs than using just gradient-based optimization or commercial tools \citep{zemax}. Exploring the design space involves changing the weights of losses for sharpness versus speed in \cref{eqn:total_loss}, creating a Pareto front of lens designs. Initialized with lenses from the Pareto front produced using only gradient-based optimization \citep{teh2024apertureaware}, our method with the same total loss finds designs that move past this front to more favorable parts of the design space. By varying both continuous and discrete lens parameters, our method produces lenses with better overall performance.

\begin{figure}[!t]
	\includegraphics{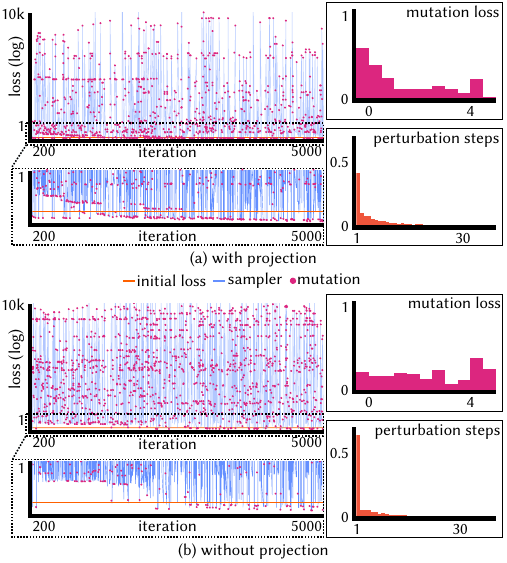}
	\caption{Two runs of our method with (top) and without (bottom) paraxial projection. We initialize both optimizations with the \qty{135}{mm} lens and run for 5000 iterations. We also plot the loss of the initial lens (orange) for comparison. Using paraxial projection results in better mutations with lower losses, and perturbation sequences with longer durations. As the reservoir saturates with good lenses, both runs more consistently find better lenses. However, paraxial projection reduces this ``warm-up'' period.}
	\label{fig:projgraph}
\end{figure}

\paragraph{Paraxial projection ablation} \Cref{fig:projgraph,tbl:projection_ablation} show results from an ablation study evaluating the utility of paraxial projection. We run our method with and without paraxial projection for 5000 iterations. The table shows that our method samples lenses that improve on the initialization much more often with paraxial projection than without. \Cref{fig:projgraph} shows the distribution of mutations during sampling for the \qty{135}{mm} lens, with and without paraxial projection. In both cases, there is a ``warm-up'' period where our method must first populate the reservoir with good lenses, before it starts consistently finding better lenses. Using paraxial projection greatly shortens this period, resulting in overall better lenses.

\begin{table}
	\centering
	\caption{Ablation study for paraxial projection. We run our method with and without paraxial projection, and report the fraction of iterations in which the sampled lenses have better objective value than the initial lens. We report results after 1000 and 5000 iterations.}
	\begin{tabularx}{\linewidth}{@{\extracolsep{\fill}} {l}*{4}{c}@{}}
		\toprule & \multicolumn{2}{c}{projection} & \multicolumn{2}{c}{no projection} \\
		\midrule
		lens & 1k & 5k & 1k & 5k \\
		\midrule
		wide-angle (\qty{28}{mm}) & 0.356 & 0.358 & 0.225 & 0.241 \\
		normal (\qty{50}{mm}) & 0.964 & 0.972 & 0.554 & 0.828 \\
		macro (\qty{105}{mm}) & 0.118 & 0.424 & 0.0 & 0.015 \\
		telephoto (\qty{135}{mm}) & 0.087 & 0.281 & 0.009 & 0.122 \\
		\bottomrule
	\end{tabularx}
	\label{tbl:projection_ablation}
\end{table}

\begin{figure}
	\includegraphics{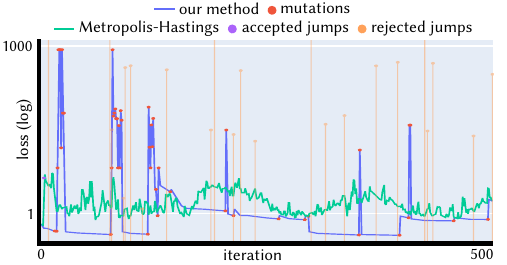}
	\caption{We compare our method with an MH-based sampler. The MH-based sampler (green) rejects most mutation proposals and can only perform small gradient steps. By contrast, our method (blue) efficiently optimizes all mutation proposals before trying another mutation.}
	\label{fig:mcmc_compare}
\end{figure}

\paragraph{Comparison with Metropolis-Hastings} In \cref{fig:mcmc_compare}, we compare our method with a reversible-jump MCMC sampler that uses the Metropolis-adjusted Langevin algorithm \citep{roberts1996exp}. As we explain in \cref{sec:mcmc}, the need to include noise in Langevin perturbations forces the MH-based sampler to use very small step sizes to ensure perturbations are accepted, slowing down optimization. The MH-based sampler also rejects most mutation proposals. By contrast, our method uses large step sizes and, as it has no rejections, attempts to optimize mutation proposals before another mutation. As a result, our method consistently finds better lenses.

\paragraph{Comparison with brute-force search} In \cref{fig:naive_comparison}, we compare our method against a baseline using brute-force search---adding or removing a singlet at every possible location in the compound lens, then optimizing with gradient descent. Inserted singlet are sampled with random thickness (mean and standard deviation \qty{1}{\milli\meter}) and curvatures (mean zero, standard deviation \qty{0.01}{\milli\meter^{-1}}). This baseline is simple to implement, but is sensitive to local minima, and becomes intractable as we increase the number of elements in the original lens or the number of elements we want to add. By contrast, our method better avoids local minima and explores more lens designs.



We use four initial lenses: a \qty{28}{mm} wide-angle lens, a \qty{50}{mm} normal lens, a \qty{105}{mm} macro lens, and a \qty{135}{mm} telephoto lens. We run our method and the baseline for \qty{40}{\min}, optimizing lenses generated from additions and removals by the baseline for an equal number of gradient iterations. In all cases, our method finds lenses with better objective values, sharpness, and speed. 

\section{Conclusion and limitations} \label{sec:limitations}

\begin{figure*}
	\includegraphics{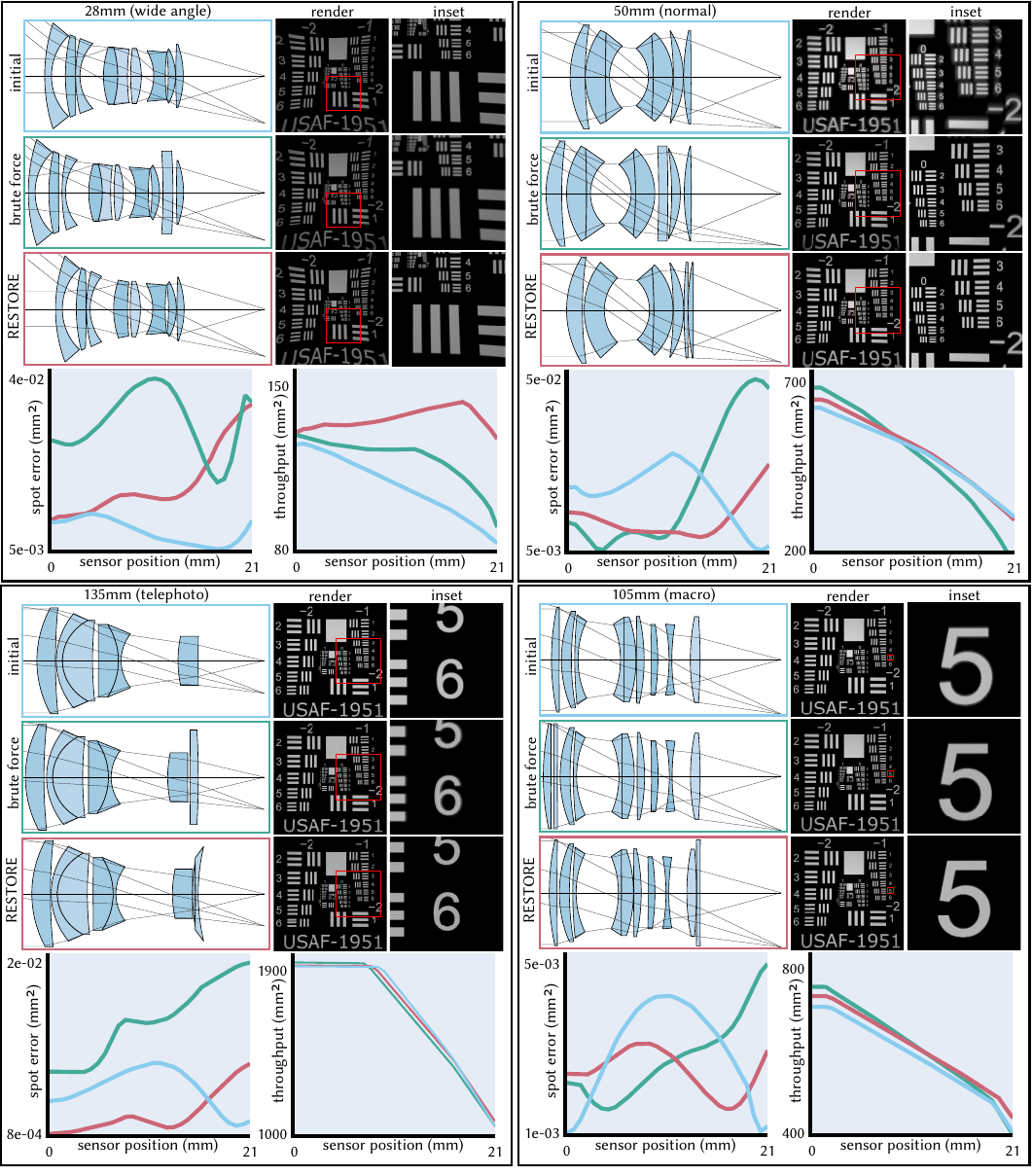}
	\vspace{-2em}
	\caption{Comparison of our method with a baseline using brute force search. The baseline adds or removes an element at every possible location in the original compound lens, then uses gradient-based optimization to improve the resulting lens. Both methods run for \qty{40}{\min}. We experiment on four different lens types: a \qty{28}{mm} wide-angle lens, a \qty{50}{mm} normal lens, \qty{105}{mm} macro lens, and a \qty{135}{mm} telephoto lens. In all cases, our method finds better lenses than the baseline for the given objective function.}
	\label{fig:naive_comparison}
\end{figure*}

We presented a method for automated design of compound lenses that, uniquely among related work, can optimize both continuous and discrete parameters of a lens design. It uses MCMC sampling to combine gradient-based optimization of continuous parameters, and tailored mutations altering the number and type of lens elements. Our method uses the \restore algorithm for effective sampling, and paraxial projection to improve mutations. It finds better lenses than using only gradient-based optimization, and expands the Pareto front possible when considering the lens speed-sharpness tradeoff. We discuss limitations that point towards future research directions.

\paragraph{Initialization dependence} Though our method explores a larger design space than using only gradient-based optimization, it is still sensitive to initialization, requiring high-quality initial designs. Sampling such designs at random is extremely unlikely, but data-driven methods \citep{cote2021autolens} provide a potential solution to this problem. Such methods can work in conjunction with ours to, e.g., prepopulate the reservoir. Alternatively, we can combine our method with curriculum learning, which has proven effective in mitigating initialization dependence for lens design tasks \citep{yang2024curriculum}.


\paragraph{Manufacturing constraints and tolerances} Though our method outputs lenses with good simulated performance, ensuring that these lenses are manufacturable and robust to manufacturing errors is essential for practicality. For example, some designs (\cref{fig:naive_comparison}) include very thin optical elements that can be either impossible to manufacture, or very unstable due to manufacturing tolerances. As a result, realizations of these designs may in practice underperform realizations of alternative designs that are worse in simulation but more reliable to manufacture. Incorporating manufacturing constraints and tolerances in the design process is possible to some extent through better engineering of the lens design objective (e.g., more strongly penalizing thin and high-curvature elements). However, given how sensitive the performance of a compound lens is to small changes in its elements (\cref{fig:projection_demo}), developing more principled solutions to these issues is an important future research direction.

\paragraph{Other design objectives} Our method optimizes objectives (\cref{eqn:total_loss}) for a single wavelength and focal length. Lens design tasks often require considering large spectral ranges (achromats and apochromats that minimize chromatic aberrations) or focal length ranges (zoom lenses that remain sharp and fast with varied focal length). Our method can be adapted to such tasks by augmenting \cref{eqn:total_loss} to average losses for different wavelengths and focal lengths, and allowing element positions to vary with focal length setting \citep[Figure 10]{teh2024apertureaware}. Gradient perturbations trivially extend to such augmented objectives, which remain differentiable. However, effective discrete optimization would require two non-trivial modifications to our method, which we leave as future research:
\begin{enumerate*}
	\item Adding mutations that change material (e.g., creation of achromatic doublets, insertion of low-dispersion singlets). 
	\item Modifying paraxial projection to account for multiple wavelengths or focal lengths.
\end{enumerate*}

\paragraph{Other geometric optical elements} Our method is limited to only spherical elements, due to our use of ray transfer matrices for paraxial projection. First-order approximations to other element types (e.g., aspherics, cylindrical lenses, or even reflective elements) exist but have lower accuracy than those for spherical elements. Thus it is unclear how effective they would be when designing lens mutations.

\paragraph{Wave optics} As we rely on geometric optics for ray tracing and paraxial projection, our method is limited to elements well approximated by geometric optics. Several recent works extend continuous lens design methods using gradient-based optimization to account for wave effects (e.g., diffraction) \citep{ho2024differentiablewaveopticsmodel, chen2023h}. Incorporating these works into our method would allow likewise extending mixed continuous-discrete lens design.

\begin{acks}
We thank Andi Wang for discussions about Restore, and Ozan Cakmakci for discussions on commercial lens design tools. This work was supported by the National Science Foundation (awards 2047341 and 2238485), the Air Force Office of Scientific Research (FA 95502410244), Alfred P. Sloan Research Fellowship FG202013153 for Ioannis Gkioulekas, and a gift from Google Research.
\end{acks}

\appendix

\section{Jacobian of paraxial projection}
\label{sec:appendix_jacobian}

We can compute the Jacobian of the solution $\params^*$ of \cref{eqn:projection} with respect to the pre-mutation parameters $\params$ analytically, using the implicit function theorem. To this end, we first write the Lagrangian of the constrained optimization problem in \cref{eqn:projection}:
\begin{equation}
	\mathcal{L}(\params^*, \params, \lambda) \coloneq \lVert \params - \params^*\rVert^2 + \lambda^\top \paren{\raytransfermat\paren{\params} - \raytransfermat\paren{\params^*}} \begin{bmatrix} 0 \\ 1 \end{bmatrix}. \label{eqn:projection_lagrange}
\end{equation}
From the method of Lagrange multipliers, the solution of \cref{eqn:projection} is a stationary point of the Lagrangian, and thus satisfies the following system of equations:
\begin{gather}
	\nabla_1 \mathcal{L} = 0, \\
	\nabla_3 \mathcal{L} = 0,
\end{gather}
where $\nabla_i$ refers to differentiation with respect to the $i$-th argument. Differentiating both sides of these equations with respect to $\params$,
\begin{gather}
	\nabla_1 \nabla_1 \mathcal{L} \frac{\ud \params^*}{\ud \params} + \nabla_2 \nabla_1 \mathcal{L} + \nabla_3 \nabla_1 \mathcal{L} \frac{\ud \lambda}{\ud \params} = 0, \label{eqn:system1}\\
	\nabla_1 \nabla_3 \mathcal{L} \frac{\ud \params^*}{\ud \params} + \nabla_2 \nabla_3 \mathcal{L} + \nabla_3 \nabla_3 \mathcal{L} \frac{\ud \lambda}{\ud \params} = 0. \label{eqn:system2}
\end{gather}
\(\nabla_1 \mathcal{L}\) and \(\nabla_2 \mathcal{L}\) have the same dimensionality, whereas \(\nabla_3 \mathcal{L}\) has dimension equal to the number of constraints (two in \cref{eqn:projection_lagrange}). We can write \cref{eqn:system1,eqn:system2} in matrix form as:
\begin{equation}
	\begin{bmatrix}
		\nabla_1 \nabla_1 \mathcal{L} & \nabla_3 \nabla_1 \mathcal{L} \\
		\nabla_1 \nabla_3 \mathcal{L} & \mathbf{0}
	\end{bmatrix}
	\begin{bmatrix}
		\frac{\ud \params^*}{\ud \params} \\
		\frac{\ud \lambda}{\ud \params}
	\end{bmatrix} =
	\begin{bmatrix}
		-\nabla_2 \nabla_1 \mathcal{L} \\
		-\nabla_2 \nabla_3 \mathcal{L}
	\end{bmatrix}.
\end{equation}
Then using \cref{eqn:projection_lagrange},
\begin{equation}
	\begin{bmatrix}
		\mathbf{I} + \lambda^\top \nabla_1 \nabla_1 g & \nabla_1 g \\
		\nabla_1 g^\top & \mathbf{0}
	\end{bmatrix}
	\begin{bmatrix}
		\frac{\ud \params^*}{\ud \params} \\
		\frac{\ud \lambda}{\ud \params}
	\end{bmatrix} =
	\begin{bmatrix}
		-\left(\mathbf{I} + \nabla_2 \nabla_1 g \right) \\
		-\nabla_2 g^\top
	\end{bmatrix},
\end{equation}
where $g$ is the function describing the constraints in \cref{eqn:projection}:
\begin{equation}
	g(\params^*, \params) \coloneq \left( \raytransfermat\paren{\params} - \raytransfermat\paren{\params^*} \right) \begin{bmatrix} 0 \\ 1 \end{bmatrix}.
\end{equation}

\bibliographystyle{ACM-Reference-Format}
\bibliography{lens_mutations.bib}

\end{document}